\documentclass[apjl]{emulateapj}

\newcommand{\PWN}{\rm{PWN}}
\newcommand{\PWNe}{\rm{PWNe}}

\newcommand{\hess}{\textit{H.E.S.S.}}

\newcommand{\dede}[2]{\frac{\partial{#1}}{\partial{#2}}}
\newcommand{\myemail}{fabio.mattana@apc.univ-paris7.fr.}

\shorttitle{Evolution of the $\gamma$- and X-ray luminosities of PWNe}
\shortauthors{Mattana et al.}

\begin{document}
\title{On the evolution of the $\gamma$- and X-ray luminosities of Pulsar Wind Nebulae}

\author{F. Mattana,\altaffilmark{1,2} 
M. Falanga,\altaffilmark{3} D. G\"otz,\altaffilmark{3} R. Terrier,\altaffilmark{1} P. Esposito,\altaffilmark{2,4,5} 
\\ A. Pellizzoni,\altaffilmark{2}  A. De Luca,\altaffilmark{2,6,5}  V. Marandon,\altaffilmark{1}
A. Goldwurm,\altaffilmark{1,3} P. A. Caraveo\altaffilmark{2} 
} 

\altaffiltext{1}{AstroParticule et Cosmologie (APC), CNRS, Universit\`e Paris 7 Denis Diderot, F-75205 Paris, France; \myemail}
\altaffiltext{2}{INAF --Istituto di Astrofisica Spaziale e Fisica Cosmica, via Bassini 15, I-20133 Milano, Italy}
\altaffiltext{3}{CEA Saclay, DSM/IRFU/Service d'Astrophysique, F-91191 Gif-sur-Yvette, France}
\altaffiltext{4}{Universit\`a degli Studi di Pavia, Dipartimento di Fisica Nucleare e Teorica, via Bassi 6, I-27100 Pavia, Italy}
\altaffiltext{5}{Istituto Nazionale di Fisica Nucleare, sezione di Pavia, via Bassi 6, I-27100 Pavia, Italy}
\altaffiltext{6}{Istituto Universitario di Studi Superiori, v.le Lungo Ticino Sforza 56, I-27100 Pavia, Italy}

\begin{abstract}
Pulsar wind nebulae are a prominent class of very high energy ($E >
0.1$ TeV) Galactic sources. Their $\gamma$-ray spectra are interpreted as due to inverse 
Compton scattering of ultrarelativistic electrons on the ambient photons, whereas
the X-ray spectra are due to synchrotron emission. We investigate the
relation between the $\gamma$- and-X-ray emission and the
pulsars' spin-down luminosity and characteristic age. We find that the distance-independent
$\gamma$- to X-ray flux ratio of the nebulae is inversely proportional to the spin-down
luminosity, ($\propto \dot{E}^{-1.9}$), while it appears proportional to the characteristic age, 
($\propto \tau_c^{2.2}$), of the parent pulsar. 
We interpret these results as due to the evolution of the electron
energy distribution and the nebular dynamics, supporting the idea
of so-called relic pulsar wind nebulae.
These empirical relations provide a new tool to classify unidentified
diffuse $\gamma$-ray sources and to estimate the spin-down
luminosity and characteristic age of rotation powered pulsars with no detected pulsation
from the X- and $\gamma-$ray properties of the associated pulsar wind nebulae. 
We apply these relations to predict the spin-down luminosity
and characteristic age of four (so far unpulsing) candidate pulsars associated to wind nebulae.
\end{abstract}


\section{Introduction}
Pulsar Wind Nebulae (PWNe) arise when the wind ejected from a rotation
powered pulsar is confined by the pressure of the surrounding
medium, be it their supernova remnant or compressed interstellar gas
\citep[see][for a review]{Gaensler06}. 
The Galactic survey performed by the \hess\ experiment
\citep[\textit{High Energy Stereoscopic System},][]{Hinton04} has
detected several \PWNe\ making them a prominent class of Very High
Energy Galactic sources \citep{HESSscanII,Gallant07,Funk07a}. In
addition to the classical investigations through radio and X-ray astronomies, Very High Energy
$\gamma$-rays (VHE $\gamma$-rays, $E > 0.1$ TeV) provide a new probe of
the physical conditions in PWNe \citep[e.g.,][]{deJager08}. 

The \PWN\ broad-band radiation is most likely due to electron-positron
pairs of the pulsar wind generated close to the magnetosphere. The wind
flow is ultrarelativistic \citep[bulk Lorentz factor $\Gamma_W \sim
  10^6$ in the Crab Nebula;][]{kc84a,kc84b}, until it experiences a
strong shock, where electrons are accelerated. After the shock,
the flow speed is sub-relativistic at the outer edge of
the \PWN. Depending on the radiation mechanisms at work,
the electrons can produce photons in different energy ranges: while synchrotron radiation
yields photons with energies up to several MeV, inverse Compton
scattering of the ambient photon field can produce
high energy photons, up to tens of TeV.  

The electrons responsible for the \PWNe\ $\gamma$-ray emission (here
after $\gamma$-ray electrons) are likely less energetic than those generating
the X-ray one (X-ray electrons), their synchrotron radiation lying at
infrared, optical, or ultraviolet frequencies. For typical nebular
magnetic field intensities ($B \sim 1$--100 $\mu$G), synchrotron
photons with energy $\sim$1 keV are produced by electrons with
Lorentz factor $\sim$0.3--3 $\times 10^8$. The Cosmic
Background Radiation, the dust-scattered light, and the starlight
provide the target photons for inverse Compton scattering, with
typical photon energies around $10^{-3}$ eV, $10^{-2}$ eV, and 1 eV,
respectively. In the Thomson regime, photons with energy $\sim$1 TeV
are produced by electrons with Lorentz factor $\sim$0.1--3 $\times
10^7$. Due to their different energies, the cooling time of the X-ray
electrons is smaller than the one of the $\gamma$-ray
electrons. Therefore, the X-ray emission traces the recent history of
the nebula, whereas the $\gamma$-ray emission traces a longer history,
possibly up to the pulsar birth. 
The different lifetime of the electrons, together with the interaction with the ambient 
medium, can lead to the significant projected angular separation, sometimes measured between
the peaks of the $\gamma$- and X-ray brightness profiles \citep[e.g. G18.0--0.7,][]{HESSJ1825b}.
Since the source of injected electrons, the
pulsar rotational energy loss rate dubbed spin-down luminosity,
decreases as time goes by, we expect a different
evolution of the $\gamma$- and X-ray luminosities, following the
particle aging and the pulsar spin-down. 

In this paper we address first the correlations between the
\PWN\ VHE $\gamma$-ray luminosities (1--30 TeV) and their X-ray
luminosities (2--10 keV) with the spin-down
luminosities, $\dot{E}$, and the characteristic ages, $\tau_c$, of
their pulsars. Next we consider the behaviour of the ratio between
the gamma and X-ray luminosity as a function of the pulsar spin-down power and age.
 These relations are discussed in the frame of an evolving
electron energy population. 

\section{Observed correlations}
In Table \ref{tab:tevpwne} we report a sample of the identified
\PWNe\ observed by the \hess\ experiment. We further included six
candidate \PWNe, selecting unidentified \hess\ diffuse sources located
near young and energetic pulsars, with $\tau_c \lesssim 100$ kyr and
$\dot{E} > 10^{35}$ erg s$^{-1}$. These parameters are defined as $\dot
E \equiv 4\pi^2 I \dot P/P^3$ and $\tau_c \equiv\ P/2\dot P$, where
$P$ is the pulsar spin period, $\dot{P}$ its derivative, and $I \equiv
10^{45}$ gm cm$^{2}$ the moment of inertia.  We calculated $\dot{E}$
and $\tau_c$ using the $P$ and $\dot{P}$ values
reported in the Australia Telescope National Facility (ATNF) pulsar catalogue\footnote{{\tt http://www.atnf.csiro.au/research/pulsar/psrcat}}
\citep{ATNF}. The $\gamma$-ray fluxes, $F_\gamma$, are derived from literature and computed in the 1--30 TeV energy band, with
statistical errors estimated with standard Montecarlo propagation technique. 
The lower energy value corresponds to the highest observed
detection threshold. The upper value of 30 TeV reduces the bias of
possible unmeasured high-energy cut-offs. The unabsorbed X-ray fluxes, $F_X$,
have been derived from literature based on X-ray imaging
observatories, and converted in the 2--10 keV energy band. The lower
energy is chosen in order to minimize the contamination by possible
thermal components due to the pulsar or supernova remnant. When it was possible to
single out the PWN from the pulsar component, only the PWN flux is reported.

\begin{deluxetable*}{lllclcll}[!h]
\tabletypesize{\scriptsize}
\tablewidth{\textwidth}
\tablecolumns{7}
\tablecaption{\label{tab:tevpwne} Properties of Wind Nebulae observed with \hess\ and associated Pulsars}
\tablehead{
\colhead{Source} & \colhead{Associated} &
\colhead{$F_{\gamma}^a$ (1--30 TeV)} &  \colhead{ $F_X^b$ (2--10 keV)} &
\colhead{ {$\tau_c$} } & \colhead{ {$\dot E$} }  & \colhead{Distance}&
\colhead{References} \\
\colhead{Name} &  \colhead{Pulsar} &
\colhead{10$^{-12}$ erg  cm$^{-2}$ s$^{-1}$} &  \colhead{erg  cm$^{-2}$ s$^{-1}$} &
\colhead{kyr} & \colhead{erg s$^{-1}$} & \colhead{kpc}
& \colhead{}
}
\startdata
Crab                         & PSR~B0531$+$21   & 80\phd\phn\      (4)\phd\phn\    (16)        &   2.10     $\times$ 10$^{-8\phn}$  &  1.2  & 4.6 $\times$ 10$^{38}$   & 1.93$^{+0.11}_{-0.11}$    & 1,2,3    \\
Vela                         & PSR~B0833$-$45   & 79\phd\phn\      (15)\phd\       (16)        &   5.39     $\times$ 10$^{-11}$     &  11   & 6.9 $\times$ 10$^{36}$   & 0.287$^{+0.019}_{-0.017}$ & 4,5,6    \\
K3 in Kookaburra             & PSR~J1420$-$6048 & 14.5             (1.6)           (2.9)       &   1.3\phn\ $\times$ 10$^{-12}$     &  13   & 1.0 $\times$ 10$^{37}$   & 5.6$^{+0.9}_{-0.8}$       & 7,8,9    \\
MSH~15--52                    & PSR~B1509$-$58   & 20.3             (1.1)           (4.1)       &   2.86     $\times$ 10$^{-11}$     &  1.6  & 1.8 $\times$ 10$^{37}$   & 5.2$^{+1.4}_{-1.4}$       & 10,11,12    \\
G18.0--0.7                    & PSR~B1823$-$13   & 61\phd\phn\      (7)\phd\phn\    (12)        &   4.4\phn\ $\times$ 10$^{-13}$     &  21   & 2.8 $\times$ 10$^{36}$   & 3.9$^{+0.4}_{-0.4}$       & 13,14,9   \\
G21.5--0.9                    & PSR~J1833$-$1034 & 2.4\phn\         (1.1)           (0.5)       &   4.0\phn\ $\times$ 10$^{-11}$     &  4.9  & 3.4 $\times$ 10$^{37}$   & 3.3$^{+0.4}_{-0.5}$       & 15,16,9  \\
AX J1838.0-0655     & PSR~J1838$-$0655   & 18.0         (2.7)           (3.6)            &   1.0\phn\ $\times$ 10$^{-12}$     &  23  & 5.5 $\times$ 10$^{36}$   & 6.6$^{+0.9}_{-0.9}$       & 17,18,19  \\
Kes 75                       & PSR~J1846$-$0258 & 2.3\phn\         (0.6)           (0.5)       &   2.27     $\times$ 10$^{-11}$     &  0.73 & 8.1 $\times$ 10$^{36}$   & 6.3$^{+1.2}_{-1.2}$      & 15,20,21  \\
HESS~J1303$-$631$^{\dagger}$ & PSR~J1301$-$6305 & 12\phd\phn\      (1.2)           (2.4)       &   6.2\phn\ $\times$ 10$^{-14}$     &  11   & 1.7 $\times$ 10$^{36}$   & 6.6$^{+1.2}_{-1.1}$       & 22,23,9  \\
HESS~J1616$-$508$^{\dagger}$ & PSR~J1617$-$5055 & 21\phd\phn\      (3)\phd\phn\    (4)         &   4.2\phn\ $\times$ 10$^{-12}$     &  8.1  & 1.6 $\times$ 10$^{37}$   & 6.7$^{+0.7}_{-0.7}$       & 17,24,9  \\
HESS~J1702$-$420$^{\dagger}$ & PSR~J1702$-$4128 & 9.1\phn\         (3.4)           (1.8)       &   6.0\phn\ $\times$ 10$^{-15}$     &  55   & 3.4 $\times$ 10$^{35}$   & 4.7$^{+0.5}_{-0.5}$       & 17,25,9  \\
HESS~J1718$-$385$^{\dagger}$ & PSR~J1718$-$3825 & 4.3\phn\         (1.3)           (0.9)       &   1.4\phn\ $\times$ 10$^{-13}$     &  90   & 1.3 $\times$ 10$^{36}$   & 3.6$^{+0.4}_{-0.4}$       & 26,27,9  \\
HESS~J1804$-$216$^{\dagger}$ & PSR~B1800$-$21   & 11.8             (1.6)           (2.4)       &   6.8\phn\ $\times$ 10$^{-14}$     &  16   & 2.2 $\times$ 10$^{36}$   & 3.8$^{+0.4}_{-0.5}$       & 17,28,9  \\
HESS~J1809$-$193$^{\dagger}$ & PSR~J1809$-$1917 & 19\phd\phn\      (4)\phd\phn\    (4)          &   2.3\phn\ $\times$ 10$^{-13}$     &  51   & 1.8 $\times$ 10$^{36}$   & 3.5$^{+0.4}_{-0.5}$       & 26,29,9
\vspace{0.1cm}
\enddata
\tablenotetext{ }{$\dagger$Candidate sources. $^a\gamma$-ray
  fluxes, statistical, and systematical errors. When not stated in the
  original papers, the systematic errors were assumed at the typical
  value of 20\% as in \citet{HESSscanII}. $^b$Errors are conservatively estimated at 20\%.}
\tablenotetext{ }{References.--
(1)  \citealt{HESSCrab};             (2) \citealt{Willingale01};    (3)  \citealt{Trimble73};    %Crab
(4)  \citealt{HESSVela};              (5) \citealt{Manzali07};        (6)  \citealt{Dodson03};   %Vela
(7)  \citealt{HESSKooka};           (8) \citealt{Ng05};               (9)  \citealt{ATNF};        %K3
(10) \citealt{HESSB1509};           (11) \citealt{Gaensler02};    (12) \citealt{Gaensler99}; %B1509 
(13) \citealt{HESSJ1825b};         (14) \citealt{Gaensler03};  %(9)  \citealt{ATNF};        %B1823
(15) \citealt{HESSG21Kes75};    (16) \citealt{Slane00};      %(9)  \citealt{ATNF};        %G21
(17) \citealt{HESSscanII};            (18) \citealt{Gotthelf08};   (19) \citealt{Davies08} %AX J1838/HESS J1837
(20) \citealt{Helfand03};             (21) \citealt{Leahy08};      %(15) \citealt{HESSG21Kes75};   %Kes75
(22) \citealt{HESSJ1303};            (23) XMM public data archive;    %(9)  \citealt{ATNF};        %J1303
 (24) \citealt{Kargaltsev08b};    %(17) \citealt{HESSscanII}; %(9)  \citealt{ATNF};       %J1617
(25) \citealt{Chang08};             %(9)  \citealt{ATNF};      (17) \citealt{HESSscanII};  %J1702
(26) \citealt{HESSJ1809_J1718}; (27) \citealt{Hinton07}; %(9)  \citealt{ATNF};                        %J1718
(28) \citealt{Kargaltsev07a}; %(17) \citealt{HESSscanII}; (9)  \citealt{ATNF};  %J1804
(29) \citealt{Kargaltsev07b}. }%(25) \citealt{HESSJ1809_J1718}; (9)  \citealt{ATNF}; %J1809
\end{deluxetable*}

\begin{figure}[!thb]
	\includegraphics[angle=0,width=0.48\textwidth]{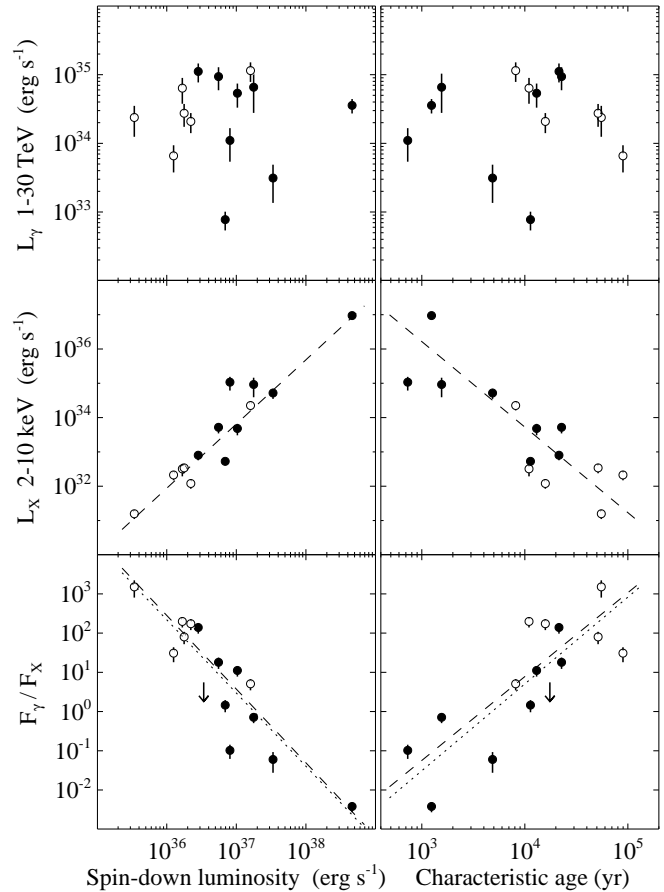} 
\caption{$\gamma$-ray luminosity, X-ray luminosity, and $\gamma$- to X-ray flux ratio
versus pulsar spin-down luminosity, $\dot{E}$ (left column),
and characteristic age, $\tau_c$ (right column). Filled and open
circles stand for identified and candidate \PWNe, respectively. The
upper-limit for the flux ratio of PSR B1706-44 \citep{HESSB1706, Romani05} 
%$F_\gamma < 1.4 \times 10^{-12}$ erg cm$^{-2}$ s$^{-1}$ >350 GeV
%$F_X = 2.52 \times 10^{-13}$ erg cm$^{-2}$ s$^{-1}$ 2-10 KeV equatorial PWN unabsorbed
is reported with an arrow. Also shown are the best-fit curves for identified
\PWNe\ (dotted lines), and for the whole sample (dashed lines). 
\label{fig:fig1}
}  
\end{figure}

We investigated the relations between the different luminosities and
the pulsar parameters, using the data collected in Table
\ref{tab:tevpwne}. The $\gamma$-ray luminosities, $L_\gamma$, do not
appear correlated with the pulsar spin-down luminosities $\dot{E}$, nor
they do with the characteristic ages $\tau_c$, as shown in
Fig. \ref{fig:fig1} (top panels). This is at variance with the observed
PWNe X-ray luminosities, for which a scaling relation is apparent with
both $\dot{E}$ and $\tau_c$ (Fig. \ref{fig:fig1}, middle panels). The
 weighted least square fit on the whole dataset yields 
\begin{equation}
\label{eq:lxedot}
\log_{10} L_X = (33.8 \pm 0.04) + (1.87 \pm 0.04) \log_{10} \dot{E}_{37}.
\end{equation}
All the uncertainties are at 1$\sigma$ level, and $\dot{E} = \dot{E}_{37} \times 10^{37}$
erg s$^{-1}$. The $L_X-\dot{E}$  scaling is known for the pulsars as
well as for the  \PWNe. This scaling was firstly noted
by \citet{Seward88}; further, \citet{Becker97} investigate a sample of 27 pulsars with \textit{ROSAT},
yielding the simple scaling $L_{X (0.1-2.4 \, \mathrm{keV})} \simeq 10^{-3} \dot{E}$.
A re-analysis was performed by \citet{Possenti02}, who studied a sample of 39 pulsars
observed by several X-ray observatories, accounting for the statistical and systematic errors. They found
$\log_{10} L_X = (-14.36 \pm 0.01) + (1.34 \pm 0.03) \log_{10} \dot{E}$, a relation harder than Eq. (\ref{eq:lxedot}).
However, they could not separate the PWN from the pulsar contribution. A better comparison can be done with the results
from \citet{Kargaltsev08a}, who recently used high-resolution \textit{Chandra} data in order to decouple the PWN and the
pulsar fluxes. Indeed, taking $\dot{E}$, $\tau_c$, and $L_{PWN}$ in the 0.5--8 keV energy band\footnote{The X-ray luminosity
reported in \citet{Kargaltsev08a} for Kes 75 was corrected according to the distance measured by \citet{Leahy08}.} from their Tables 1 and 2,
we obtained as fitted values
$\log_{10} L_{X (0.5-8 \, \mathrm{keV})} = (34.02 \pm 0.05) + (1.46 \pm 0.04) \log_{10} \dot{E}_{37}$ for their whole sample, and
$\log_{10} L_{X (0.5-8 \, \mathrm{keV})} = (34.26 \pm 0.03) + (1.87 \pm 0.01) \log_{10} \dot{E}_{37}$ restricting the fit
only to the sources also present in our sample.
The latter is compatible in the terms of slope with Eq. (\ref{eq:lxedot}), and the slight difference in normalization can
be due to the different energy band.

X-ray sources of our whole dataset also show a dependence of
$L_X$ on $\tau_c$, with a best-fit relation 
\begin{equation}
\label{eq:lxtau}
\log_{10} L_X = (33.7 \pm 0.04) - (2.49 \pm 0.06) \log_{10} \tau_4,
\end{equation}
where $\tau_c$ is in units of years. The $L_X-\tau_c$ scaling was already noted by
\citet{Becker97} and \citet{Possenti02}. Also in this case we compared our fit to the one derived
using the whole \citet{Kargaltsev08a} dataset, which results in
$\log_{10} L_{X (0.5-8 \, \mathrm{keV})} = (34.29 \pm 0.01) - (2.03 \pm 0.01) \log_{10} \tau_4$ for their whole sample, and
$\log_{10} L_{X (0.5-8 \, \mathrm{keV})} = (34.23 \pm 0.02) - (2.60 \pm 0.02) \log_{10} \tau_4$ restricting the fit
only to the sources also present in our sample.

\begin{deluxetable*}{lccrr}
\tabletypesize{\footnotesize}
\tablewidth{0.7\textwidth}
\tablecolumns{5}
\tablecaption{\label{tab:new} PWNe hosting a neutron star without detected pulsations} 
\tablehead{
\colhead{Source} & \colhead{$F_{\gamma}$ (1--30 TeV)} & \colhead{
  $F_X$ (2--10 keV)} & \colhead{ {$\tau_c^\ast$} } & \colhead{ {$\dot{E}^\ast$} } \\
\colhead{Name} & \colhead{erg  cm$^{-2}$ s$^{-1}$} &
\colhead{erg  cm$^{-2}$ s$^{-1}$} & \colhead{kyr} & \colhead{erg s$^{-1}$}
}
\startdata
G313.+0.1 Rabbit                           &   $1.0 \times 10^{-11}$    & $7.3 \times 10^{-12}$ & $\sim$6   & $\sim1.5 \times 10^{37}$ \\
G0.9+0.1                                        &   $3.3 \times 10^{-12}$    & $5.8 \times 10^{-12}$ & $\sim$4    & $\sim2 \times 10^{37}$ \\
G12.8-0.0$^{\dagger}$                   &   $1.3 \times 10^{-11}$    & $9.2 \times 10^{-12}$ & $\sim$6    & $\sim1.5 \times 10^{37}$ \\
HESS~J1640--465$^{\dagger}$      &   $9.3 \times 10^{-12}$    & $1.0 \times 10^{-12}$ & $\sim$13  & $\sim5 \times 10^{36}$ \\
\vspace{-0.3cm}
\enddata
%\tablenotetext{a}{pinuccio.}
\tablenotetext{ }{$^\dagger$Candidate sources. $^\ast$Predicted values. References.-- HESS~J1418/G313.+0.1 Rabbit: \citet{HESSKooka}, \citet{Ng05}; HESS~J1747--281/G0.9+0.1: \citet{HESSG09}, \citet{Porquet03}; HESS~J1813--178/G12.8--0.0: \citet{HESSscanII}, \citet{Helfand07}; HESS~J1640--465/G338.3--0.0: \citet{HESSscanII}, \citet{Funk07b}.}
\end{deluxetable*}

The lower panels of Fig. \ref{fig:fig1} refer to the $\gamma$- to X-ray flux ratio
 $F_\gamma/F_X$. There is a clear anticorrelation between $F_\gamma/F_X$ and $\dot{E}$, spanning
over four decades in $\dot{E}$  and seven decades in $F_\gamma/F_X$
(Fig. \ref{fig:fig1}, bottom left panel). Considering only the identified \PWNe,
the correlation coefficient is $r = -0.7\pm 0.2$; including also the
candidate sources, the anticorrelation is more significant, with $r =
-0.84\pm 0.09$. The best-fit including only the identified sources yields  
\begin{equation}
\label{eq:fgammafx_edot1}
\log_{10} F_\gamma/F_X = (0.47 \pm 0.05) - (1.87 \pm 0.07) \log_{10} \dot{E}_{37}.
\end{equation}
For all the data points, it results
\begin{equation}
\label{eq:fgammafx_edot2}
\log_{10} F_\gamma/F_X = (0.57 \pm 0.04) - (1.88 \pm 0.05) \log_{10} \dot{E}_{37},
\end{equation}
compatible within the errors with the relation obtained using only the identified sources.

The $\gamma$- to X-ray flux ratio is also found to correlate with the
characteristic age $\tau_c$ (Fig. \ref{fig:fig1}, bottom right panel), with
a correlation coefficient $r = 0.7 \pm 0.2$ for identified \PWNe\ only, and
$r = 0.75 \pm 0.13$ for the whole sample. The ordinary weighted least
square fit only for the identified \PWNe\ yields 
\begin{equation}
\label{eq:fgammafx_tau1}
\log_{10} F_\gamma/F_X = (0.70 \pm 0.06) + (2.21 \pm 0.09) \log_{10} \tau_4,
\end{equation}
and for all the data points
\begin{equation}
\label{eq:fgammafx_tau2}
\log_{10} F_\gamma/F_X = (0.89 \pm 0.04) + (2.14 \pm 0.07) \log_{10} \tau_4.
\end{equation}
One should note that these correlations are based on 8 identified sources,
and are consistent when the 6 candidate sources are considered.
More $\gamma$-ray detections may improve their significance.

\section{Discussion}
\label{sec:model}
We found the $\gamma$- to X- ray luminosity ratio $L_\gamma/L_X =
F_\gamma/F_X$ to be anticorrelated with the spin-down luminosity
$\dot{E}$ and correlated with the characteristic age $\tau_c$. Formally,
such dependencies are driven by the scaling law of the X-ray
luminosity $L_X$, which increases with $\dot{E}$ and decreases with
$\tau_c$, since the values of $L_\gamma$ were found uncorrelated with
the pulsar parameters. However, the $F_\gamma/F_X$ is a
distant-independent indicator which relates two electron populations,
differing by energy and age. An evolution in the \PWN\ broad-band
spectrum is pointed out by Eq. (\ref{eq:fgammafx_tau1}), which implies
$L_\gamma > L_X$ after $\sim$5 kyr  from pulsar birth. Therefore,
the $\gamma$-ray emission remains efficient around
$L_\gamma \sim 10^{33}$--10$^{35}$ erg s$^{-1}$, while the X-ray luminosity
decreases by a factor $\sim$10$^6$ in 10$^5$ yr following the pulsar
spin-down.

Such a broad-band spectral evolution can be expected on the basis of
the \PWNe\ leptonic model \citep{kc84b, Chevalier00}. In a PWN, the
source of the injected electrons is the pulsar spin-down luminosity,
$\dot{E}$. The total injection rate of the electrons can be written: 
\begin{equation}
\dot{N} = \frac{\dot{E}}{\Gamma_W \, m_e c^2 \, (1 + \sigma)},
\end{equation}
where the magnetization parameter $\sigma$ sets the fraction of the spin-down luminosity converted in kinetic energy of the wind. The whole spin-down luminosity is converted in particle kinetic energy for $\sigma \ll 1$, as for the Crab Nebula \citep{kc84a,kc84b}. For sake of simplicity, we assume a constant wind Lorentz factor $\Gamma_W$ upstream the shock.  $\dot{E}$ decreases in time as \cite[e.g.,][]{Pacini73}
\begin{equation}
\label{eq:varying-edot2}
\dot{E}(t) = \frac{\dot{E}_0}{ \left( 1 + t/t_{dec} \right)^\beta},
\end{equation}
where $\dot{E}_0 \sim 10^{38}$--10$^{40}$ erg s$^{-1}$ is the
spin-down luminosity at the pulsar birth, $t_{dec} \sim 100$--1000 yr is a
characteristic decay time, $t$ is the time elapsed since pulsar birth
($t_0 = 0$), and $\beta = (n+1)/(n-1)$, where $n$ is the braking index.
In the following, we assume a  pure dipolar magnetic field torque, i.e. $n = 3$.
As the braking indices inferred from the measurement of the period and its derivatives
are significantly smaller than 3 \citep{Livingstone07}, we dealt with a generic $n$ (see App. \ref{app:a}),
and found that the results derived from Eq. (\ref{eq:varying-edot2}) are unaffected by the choice of $n$.

Since it depends on $\dot{E}$, also the particle injection rate $\dot{N}$
decreases in time. Therefore, the total number of particles  
\begin{equation}
\label{eq:N}
N \propto \int_0^t \dot{E}(t') \, dt' = \dot{E}_0 \, t_{dec} \left(
\frac{t}{t+t_{dec}} \right), 
\end{equation}
reaches a constant value $N \propto \dot{E}_0 \, t_{dec}$ for $t \gg
t_{dec}$, and the particle supply by the pulsar becomes negligible. 

The electron energy distribution $n(E,t)$ accounting for particle
injection and radiative losses evolves according to the kinetic
equation \citep[e.g.,][]{Ginzburg64,Blumenthal70}:  
\begin{equation}
\label{eq:kineticeq}
\dede{n}{t} = \dede{}{E} \left(n P\right) + Q,
\end{equation}
where  $Q = Q(E,t)$ is the particle distribution injected per unit time, and $P = P(E,t)$
is the radiated power per particle with energy $E$. The normalization
of $n(E,t)$ is set by $N$ via the injection rate: $\dot{N}(t) = \int
Q(E,t) dE$. 

At energies for which the radiative losses are negligible, the number
of particles $n(E,t)$ with energy $E$ at time $t$ has the same profile
of the injected distribution $Q(E)$ with a normalization set by
$N$. Therefore, 
\begin{equation}
\label{eq:n_uncooled}
n_{u}(E,t) \propto \int_0^t \dot{E}(t') \, dt' = \dot{E}_0 \, t_{dec}
\left( \frac{t}{t+t_{dec}} \right), 
\end{equation}
where $u$ stands for {\em uncooled}.  As in Eq. (\ref{eq:N}), a constant 
value $n_{u}(E,t) \propto \dot{E}_0 \, t_{dec}$ is reached for $t \gg t_{dec}$. 

The effect of the radiative losses is to limit the accumulation of
particles at a given energy. After an energy-dependent cooling time
$t_c(E)$, the particles with initial energy $E$ have radiated a
significant fraction of their energy \citep{Chevalier00}. Accounting
for pitch-angle averaged synchrotron and inverse Compton in the Thomson regime energy losses, 
the cooling time can be written as 
\begin{equation}
\label{eq:coolingtime}
t_c (E) = \frac{9 \, m_e^3 c^5}{4 \, (1+\xi) \, e^4 \, \gamma_E  \, B^2} \simeq
24.5 \, (1+\xi)^{-1}\, \gamma_7^{-1} \, B_5^{-2} \quad \textrm{kyr}, 
\end{equation}
where  $\gamma_E= E/(m_e c^2)$ is the particle Lorentz factor, and $\xi
= U_{ph}/U_B$, with $U_{ph}$ and $U_B$ the photon field and
magnetic field energy densities, respectively ($\gamma_E = \gamma_7 \times 10^7$, $B = B_5 \times 10^{-5}$ G). 
When the photon field is provided by the Cosmic Background Radiation ($U_{ph} = 0.26$ eV cm$^{-3}$), the synchrotron
radiation is the main cooling process ($\xi < 1$) if $B > 3$
$\mu$G. This condition is generally fulfilled in \PWNe\ as the
equipartition magnetic field intensity ranges in $B \sim 1 - 100$
$\mu$G.\footnote{In radiation-dominated environment, like the Galactic
  Center, the inverse Compton can contribute to the cooling. In this
  case, the Klein-Nishina regime should be taken into account
  \citep{Manolakou07}.} Eq. (\ref{eq:coolingtime}) shows that
the cooling time of $\gamma$-ray radiating particles, $t_{c\gamma}$,
is one order of magnitude longer than that of the X-ray radiating
particles, $t_{cX}$, e.g., for $B = 10$ $\mu$G, $t_{c\gamma} \sim
8$--250 kyr, and $t_{cX} \sim 0.8$--8 kyr. By comparing $t_{c\gamma}$
and $t_{cX}$ with the average characteristic ages of pulsars in TeV
\PWNe, the $\gamma$-radiation is produced by long-lived electrons
tracing the time-integrated evolution of the nebula, even up to the
pulsar birth, whereas the X-ray emission is generated by younger
electrons, injected in the last thousands of years. 

Only the particles injected since the last $t_c(E)$ years will contribute to
$n(E,t)$. Eq. (\ref{eq:n_uncooled}) is accordingly modified: 
\begin{equation}
\label{eq:n_cooled}
n_{c}(E,t) \propto \int_{t-t_c}^{t} \dot{E}(t') \, dt' =
\frac{\dot{E}_0 \, t_{dec}^2 \, t_c}{(t - t_c + t_{dec}) \, (t+t_{dec})}, 
\end{equation}
where $c$ stands for {\em cooled}. This implies $n_c(E,t) \propto
\dot{E}_0 \, t_{dec}^2  \, t_c t^{-2}$ for $t \gg {\rm
  max}(t_c,\,t_{dec})$, and hence $n_c(E,t) \propto \dot{E} \, t_c$
using Eq. (\ref{eq:varying-edot2}). 

\section{Conclusions}
\begin{figure}[htb]
%	\centering 
\includegraphics[angle=0,width=0.48\textwidth]{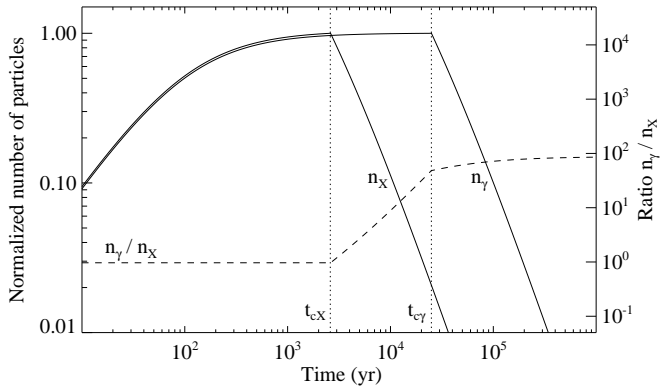} 
\caption{Time evolution of the number of particles radiating in VHE 
$\gamma$-rays, $n_\gamma$, and in X-rays, $n_X$ (solid lines), and of their ratio (dashed line). 
Pulsar birth is at $t = 0$. Initial conditions for the pulsar spin-down luminosity
are $\dot{E}_0 = 10^{39}$ erg s$^{-1}$ and $t_{dec}$ = 100 yr. Both curves are
normalized to their maximum value. After the initial rise, both 
particle populations reach a plateau. The fall begins at t
greater than the cooling time, which is  assumed to be: $t_{cX}$ = 2.6 kyr for X-rays,
$t_{c\gamma}$ = 25 kyr for $\gamma$-rays (for a magnetic field intensity $B = 10 \, \mu$G, and
a Lorentz factor of $\gamma$-ray radiating electrons $\gamma = 10^7$).
}  
\label{fig:fig2} 
\end{figure}

Eqs. (\ref{eq:n_uncooled}) and (\ref{eq:n_cooled}) describe the
time evolution of a particle populations in two regimes, uncooled and
cooled. Such an evolution is exemplified in Fig. \ref{fig:fig2} for
the populations of particles producing $\gamma$-rays, $n_\gamma$, and
X-rays, $n_X$. After the initial rise, both the particle populations
reach a plateau ($t > t_{dec}$). The decrease begins when the
evolution time is greater than the cooling time. As in general $t_{cX}
< t_{c\gamma}$, the X-ray emission fades long before the $\gamma$-ray
one. 

As the characteristic ages of the pulsars powering a VHE $\gamma$-ray
\PWN\ are in the range 1--20 kyr, likely $t_{cX} < \tau_c <
t_{c\gamma}$. Accordingly, the population of the X-ray electrons,
$n_X$, is likely to be in the cooling regime, i.e., it decreases. The
scaling laws $n_X \propto \tau_c^{-2}$ and $n_X \propto \dot{E}$ of Eq. (\ref{eq:n_cooled}) support the
trend observed in the data, see Eq. (\ref{eq:lxedot}). At variance,
the population of the $\gamma$-ray electrons, $n_\gamma$, is in the
uncooled regime, the asymptotic limit of Eq.~(\ref{eq:n_uncooled});
this explains the lack of correlation between $\gamma$-ray luminosity
$L_\gamma$ and $\dot{E}$. Finally, Eqs.~(\ref{eq:n_uncooled}) and
(\ref{eq:n_cooled}) for $t_{cX} < \tau_c <  t_{c\gamma}$ imply 
a ratio  $n_\gamma/n_X \propto t^{2} \propto \dot{E}^{-1}$. Since the luminosities
are roughly proportional to the population of radiating particles,
finally one gets 
\[
L_\gamma/L_X \propto t^{2} \propto \dot{E}^{-1},
\]
to compare with the best-fit empirical relations $L_\gamma/L_X \propto
\tau_c^{2.2}$ and $L_\gamma/L_X \propto \dot{E}^{-1.9}$, see
Eqs. (\ref{eq:fgammafx_tau1}) and (\ref{eq:fgammafx_edot1}). Though
the outlined model does not correctly predict the slopes, not surprisingly in being simplified,
it highlights the concurrent roles of the evolving pulsar injection and of the radiative losses
in producing the observed trends. 

The scattering around the relations for $F_\gamma/F_X$ reflects the
lack of correlation between $L_\gamma$ and $\dot{E}$. Environmental
factors can affect the $\gamma$-ray luminosities, like the local
energy density of the ambient photon field, or the interaction with
the surrounding medium causing an enhancement in the magnetic
field.  Also, unmeasured pulsar properties such as the magnetic field,
its orientation with respect to the spin axis, and the initial spin period
might affect the pulsar wind properties.

We stress that the relations presented here are derived under several assumptions, the most important of which being that 
X-ray and $\gamma$-ray emitting particles are in different cooling regimes, cooled for X-rays and uncooled for $\gamma$-rays.
However, the Lorentz factors ranges of the two populations get closer, and they can even overlap, if the nebular magnetic field is very high,
on the order of $B$=170 $\mu$G. On the other hand, in the case of a young nebula with a very low magnetic field, the X-ray electrons may
not have reached the cooling regime, leaving the $\gamma$-ray production to the low-energy freshly injected electrons.
Hence, PWNe with a very weak magnetic field, like 3C 58 \citep{Slane08}, or possibly with a unusually strong one, as reported lately by \citet{Arzoumanian08} for DA 495, could represent outliers to our derived relations. These regimes can be properly taken into account through numerical
modelling of the kinetic equation (Eq. [\ref{eq:kineticeq}]). Another important assumption is a uniform and constant
magnetic field: indeed high resolution imaging observations of several PWNe show 
a dynamical and structured nebular morphology \citep{Gaensler06}.
The evolution of the average magnetic field is complicated by the interaction with the supernova ejecta,
which is expected to occur after a few thousands of years since pulsar birth, causing global oscillations of
the magnetic field intensity \citep[e.g.,][]{Bucciantini03}.
One should note that the cooling time is not well defined if it is comparable to or longer than the time scale
of variation of the magnetic field. The cases of patchy or evolving magnetic field are further sources of
scattering around our relations.

Given the limitations discussed above, the empirical relations in Eqs. (\ref{eq:fgammafx_edot1}) and
(\ref{eq:fgammafx_tau1}) provide a new tool to estimate the spin-down
luminosity and characteristic age of a pulsar lacking detected pulsation
from the $\gamma$- and X-ray  properties of the associated \PWN. For
the four candidate pulsars in Table 2, we used $F_\gamma/F_X$ to predict
$\dot{E}$ and $\tau_c$. Taking into account the average scattering (average absolute ratio) around the
best fit relations, Eqs. (\ref{eq:fgammafx_edot1}) and (\ref{eq:fgammafx_tau1}),
one should expect an uncertainty of a factor $\sim$2.5 for $\dot{E}$
and $\sim$2.3 for $\tau_c$ considering only the eight identified sources.
On the other hand, considering Eqs. (\ref{eq:fgammafx_edot2}) and (\ref{eq:fgammafx_tau2}),
and including also the candidate sources, the uncertainties are $\sim$2.2 for $\dot{E}$
and $\sim$2.6 for $\tau_c$.

The correlations for $F_\gamma/F_X$  hold also after including
the candidate sources. The pulsars possibly associated to the candidate sources are 
mostly older Vela-like pulsars, with $8 \times 10^3$ yr $<$ $\tau_c$ $<$ $9 \times 10^4$ yr,
and  $3.4 \times 10^{35}$ erg s$^{-1}$ $<$ $\dot{E}$ $<$
$1.6 \times 10^{37}$ erg s$^{-1}$. Due to the pulsar ages, the
electrons had the time to be advected far from the pulsar, producing the
offset between the $\gamma$-ray emission centroid and the pulsar
position, the process leading to the so-called relic \PWNe\ \citep{deJager08}. 
The fact that all the confirmed associations contain younger pulsars is hence not surprising, as the
positional coincidence is one of the main identification criteria. 
If the identification of candidate sources with relic \PWNe\ is
confirmed, the $\gamma$-ray luminosity would persist up to 10$^{5}$
yr, with remarkable time-integrated energy channeled in radiation ($\sim 3 \times 10^{45}$--$3\times 10^{47}$ erg). 

 \acknowledgments
FM, MF, and DG acknowledge the French Space Agency (CNES) for financial support.
FM is also grateful for support from the Moscow St. NGO.
We wish to thank the referee, P. Slane, for his very constructive comments and suggestions that helped to improve the manuscript.

\begin{appendix}
\section{Particle population injected by a pulsar with generic braking index}
\label{app:a}

By adopting Eq. (\ref{eq:varying-edot2}) for a generic braking index $n > 1$,  Eqs. (\ref{eq:n_uncooled}) and (\ref{eq:n_cooled}) are so modified:
\begin{equation}
\label{eq:An_uncooled}
n_{u} (E,t) \propto \int_0^t \dot{E}(t') \, dt' = \frac{\dot{E}_0 \, t_{dec}}{\beta-1}\left[ 1 -  \left(\frac{t_{dec}}{t+t_{dec}} \right)^{\beta -1} \right], 
\end{equation}
and
\begin{equation}
\label{eq:An_cooled}
n_{c}(E,t) \propto \int_{t-t_c}^{t} \dot{E}(t') \, dt' = \frac{\dot{E}_0 \, t_{dec}^\beta}{\beta-1} \; (t_{dec} + t)^{1 - \beta} \; \left[ \left(1 - \frac{t_c}{t_{dec} + t} \right)^{1-\beta} -1 \right].
\end{equation}
For $t \gg t_{dec}$ Eq. (\ref{eq:An_uncooled}) yields $n_{u} \propto \dot{E}_0 \, t_{dec} / (\beta -1)$, while for  $t \gg {\rm
  max}(t_c,\,t_{dec})$ Eq. (\ref{eq:An_cooled}) yields  $n_{c} \propto \dot{E}(t) \, t_{c}$. As in the case of the dipolar magnetic braking, in the uncooled regime most of the radiating particles has been injected in the early phases, whereas in the cooled regime the particle population is limited by the cooling time and follows more closely the current spin-down rate.
\end{appendix}
%\clearpage

\bibliographystyle{apj}

\begin{thebibliography}

\bibitem[Aharonian et al.(2005a)]{HESSB1706}
Aharonian, F., et al. (HESS Collaboration)\ 2005a, \aap, 432, L9 

\bibitem[Aharonian et al.(2005b)]{HESSG09}
Aharonian, F., et al. (HESS Collaboration)\ 2005b, \aap, 432, L25 

%\bibitem[Aharonian et al.(2005)]{2005Sci...307.1938A}
%Aharonian, F., et al. (HESS Collaboration)\ 2005, Science, 307, 1938 

\bibitem[Aharonian et al.(2005c)]{HESSB1509}
Aharonian, F., et al. (HESS Collaboration)\ 2005c, \aap, 435, L17 

\bibitem[Aharonian et al.(2005d)]{HESSJ1303}
Aharonian, F., et al. (HESS Collaboration)\ 2005d, \aap, 439, 1013 

%HESS J1825 first paper
%\bibitem[Aharonian et 
%al.(2005)]{2005A&A...442L..25A} Aharonian, F.~A., et al. (HESS Collaboration)\ 2005, \aap, 442, L25 

\bibitem[Aharonian et al.(2006a)]{HESSVela}
Aharonian, F., et al. (HESS Collaboration)\ 2006a, \aap, 448, L43 

\bibitem[Aharonian et al.(2006b)]{HESSKooka}
Aharonian, F., et al. (HESS Collaboration)\ 2006b, \aap, 456, 245 

\bibitem[Aharonian et al.(2006c)]{HESSCrab}
Aharonian, F., et al. (HESS Collaboration)\ 2006c, \aap, 457, 899 

\bibitem[Aharonian et al.(2006d)]{HESSJ1825b}
Aharonian, F., et al. (HESS Collaboration)\ 2006d, \aap, 460, 365 

\bibitem[Aharonian et al.(2006e)]{HESSscanII}
Aharonian, F., et al. (HESS Collaboration)\ 2006e, \apj, 636, 777 

\bibitem[Aharonian et al.(2007)]{HESSJ1809_J1718}
Aharonian, F., et al. (HESS Collaboration)\ 2007, \aap, 472, 489 

\bibitem[Aharonian et al.(2008)]{HESSJ1913}
Aharonian, F., et al. (HESS Collaboration)\ 2008, \aap, 484, 435 

\bibitem[Arzoumanian et al.(2008)]{Arzoumanian08}
 Arzoumanian, Z., Safi-Harb, S., Landecker, T.~L., Kothes, R., 
\& Camilo, F.\ 2008,  \apj, in press (astro-ph/0806.3766)

\bibitem[Becker \& Tr\"umper(1997)]{Becker97}
Becker, W., \& Tr\"umper, J.\ 1997, \aap, 326, 682

\bibitem[Bucciantini et al.(2003)]{Bucciantini03}
Bucciantini, N., Blondin, J.~M., Del Zanna, L., \& Amato, E.\ 2003, \aap, 405, 617 \bibitem[Blumenthal \& Gould (1970)]{Blumenthal70}
{{Blumenthal}, G.~R. \& {Gould}, R.~J.} 1970, Rev. Mod. Phys., 42, 237

% J1702, ApJ accepted
\bibitem[Chang et al.(2008)]{Chang08}
Chang, C., Konopelko, A., \& Cui, W.\ 2008, \apj, 682, 1177

\bibitem[Chevalier(2000)]{Chevalier00}
{{Chevalier}, R.~A.} 2000, \apjl, 539, L45

\bibitem[Davies et al.(2008)]{Davies08}
Davies, B., Figer, D.~F., Law, C.~J., Kudritzki, R.-P., Najarro, F., Herrero, A., 
\& MacKenty, J.~W.\ 2008, \apj, 676, 1016

\bibitem[Djannati-Ata{\"i} et al.(2007)]{HESSG21Kes75}
{{Djannati-Ata{\"i}}, A., {de Jager}, O.~C.,  {Terrier}, R.,
  {Gallant}, Y.~A., {Hoppe}, S.} 2007, in Proceedings of the 30th ICRC
(Merida, Mexico), in press, (astro-ph/0710.2247) 

\bibitem[Dodson et al.(2003)]{Dodson03}
Dodson, R., Legge, D., Reynolds, J.~E., \& McCulloch, P.~M.\ 2003, \apj, 596, 1137  

%\bibitem[Esposito(2003)]{EspositoPC}
%Esposito, P., private communication, based on the XMM public data archive 

\bibitem[Funk(2007)]{Funk07a}
Funk, S.\ 2007, \apss, 309, 11 

\bibitem[Funk et al.(2007)]{Funk07b}
Funk, S., Hinton, J.~A., P{\"u}hlhofer, G., Aharonian, F.~A., Hofmann, W., Reimer, O., \& Wagner, S.\ 2007, \apj, 662, 517

\bibitem[Gaensler et al.(1999)]{Gaensler99}
Gaensler, B.~M., Brazier, K.~T.~S., Manchester, R.~N., Johnston, S., \& Green, A.~J.\ 1999, \mnras, 305, 724 

\bibitem[Gaensler et al.(2002)]{Gaensler02}
Gaensler, B.~M., Arons, J., Kaspi, V.~M., Pivovaroff, M.~J., Kawai, N.,  \& Tamura, K.\ 2002, \apj, 569, 87

\bibitem[Gaensler et al.(2003)]{Gaensler03}
Gaensler, B.~M., Schulz, N.~S., Kaspi, V.~M., Pivovaroff, M.~J., \& Becker, W.~E.\ 2003, \apj, 588, 441 

\bibitem[Gaensler \& Slane(2006)]{Gaensler06} 
Gaensler, B.~M., \& Slane, P.~O.\ 2006, \araa, 44, 17

\bibitem[Gallant(2007)]{Gallant07}
Gallant, Y.~A.\ 2007, \apss, 309, 197 

\bibitem[Ginzburg \& Syrovatskii(1964)]{Ginzburg64} 
Ginzburg, V.~L., \& Syrovatskii, S.~I.\ 1964, The Origin of Cosmic Rays (New York: Macmillan)  

%\bibitem[Gotthelf(2003)]{Gotthelf03}
%Gotthelf, E.~V.\ 2003, \apj, 591,361 

%AX J1838
\bibitem[Gotthelf \& Halpern(2008)]{Gotthelf08}
Gotthelf, E.~V., \& Halpern, J.~P.\ 2008, \apj, 681, 515

%\bibitem[\protect\citeauthoryear{Gould}{1965}]{Gould65}
%{{Gould}, R.~J.} 1965, PRL, 15, 577

\bibitem[Helfand et al.(2003)]{Helfand03}
Helfand, D.~J., Collins, B.~F., \& Gotthelf, E.~V.\ 2003, \apj, 582, 783 

\bibitem[Helfand et al.(2007)]{Helfand07}
Helfand, D.~J., Gotthelf, E.~V., Halpern, J.~P., Camilo, F., Semler, D.~R., Becker, R.~H., \& White, R.~L.\ 2007, \apj, 665, 1297 

\bibitem[Hinton(2004)]{Hinton04} 
Hinton, J.~A.\ 2004, NewA Rev., 48, 331

\bibitem[Hinton et al.(2007)]{Hinton07}
Hinton, J.~A., Funk, S., Carrigan, S., Gallant, Y.~A., de Jager, O.~C., Kosack, K., Lemi{\`e}re, A., P\"uhlhofer, G.\ 2007, \aap, 476, L25 

%B1800
\bibitem[Kargaltsev et al.(2007)]{Kargaltsev07a} Kargaltsev, O., 
Pavlov, G.~G., \& Garmire, G.~P.\ 2007, \apj, 660, 1413

%J1809
\bibitem[Kargaltsev  \& Pavlov(2007)]{Kargaltsev07b}
Kargaltsev, O., \& Pavlov, G.~G.\ 2007, \apj, 670, 655 

%Lx - Edot relation
\bibitem[Kargaltsev \& Pavlov(2008)]{Kargaltsev08a}
Kargaltsev, O., \& Pavlov, G.~G.\ 2008, in AIP Conf. Proc. 983, 171
%in AIP Conf. Proc. 983, 40 Years of Pulsars: Millisecond Pulsars, 
%Magnetars, and More, ed. C. Bassa, Z. Wang, A. Cumming, & V. M. Kaspi 
%(New York: AIP), 171

%1617
\bibitem[Kargaltsev et al.(2008)]{Kargaltsev08b}
Kargaltsev, O., Pavlov, G.~G., \& Wong, J.~A.\ 2008, \apj, submitted (astro-ph/0805.1041) 

\bibitem[Kennel \& Coroniti(1984a)]{kc84a}
Kennel, C.~F., \& Coroniti, F.~V.\ 1984a, \apj, 283, 694 

\bibitem[Kennel \& Coroniti(1984b)]{kc84b}
Kennel, C.~F., \& Coroniti, F.~V.\ 1984b, \apj, 283, 710 

\bibitem[\protect\citeauthoryear{de Jager \& Djannati-Ata{\"i}}{2008}]{deJager08} 
{{de Jager}, O.~C. and {Djannati-Ata{\"i}}, A.} 2008, (astro-ph/0803.0116)

\bibitem[Leahy \& Tian(2008)]{Leahy08}
Leahy, D.~A., \& Tian, W.~W.\ 2008, \aap, 480, L25

\bibitem[Livingstone et al.(2007)]{Livingstone07}
Livingstone, M.~A., Kaspi, V.~M., Gavriil, F.~P., Manchester, R.~N., Gotthelf, E.~V.~G., \& Kuiper, L.\ 2007, \apss, 308, 317 

\bibitem[\protect\citeauthoryear{Manchester et al.}{2005}]{ATNF}
{{Manchester}, R.~N. and {Hobbs}, G.~B. and {Teoh}, A. and {Hobbs}, M.} 2005, \aj, 129, 1993

\bibitem[\protect\citeauthoryear{Manolakou et al.}{2007}]{Manolakou07}
{{Manolakou}, K., {Horns}, D. \& {Kirk}, J.~G.} 2007, \aap, 474, 689

\bibitem[Manzali et al.(2007)]{Manzali07}
Manzali, A., De Luca, A., \& Caraveo, P.~A.\ 2007, \apj, 669, 570

\bibitem[Ng et al.(2005)]{Ng05}
Ng, C.-Y., Roberts, M.~S.~E., \& Romani, R.~W.\ 2005, \apj, 627, 904

\bibitem[Porquet et al.(2003)]{Porquet03}
Porquet, D., Decourchelle, A., \& Warwick, R.~S.\ 2003, \aap, 401, 197 

\bibitem[Pacini \& Salvati(1973)]{Pacini73}
Pacini, F., \& Salvati, M.\ 1973, \apj, 186, 249 

%%ref piu' nuova per B1823, non usata
%\bibitem[Pavlov et al.(2008)]{Pavlov08}
%Pavlov, G.~G., Kargaltsev, O., \& Brisken, W.~F.\ 2008, \apj, 675, 683

\bibitem[Possenti et al.(2002)]{Possenti02}
Possenti, A., Cerutti, R., Colpi, M., \& Mereghetti, S.\ 2002, \aap, 387, 993 

\bibitem[Romani et al.(2005)]{Romani05}
Romani, R.~W., Ng, C.-Y., Dodson, R., \& Brisken, W.\ 2005, \apj, 631, 480

\bibitem[Rybicki \& Lightman(1979)]{Rybicki79}
{{Rybicki}, G.~B., \& {Lightman}, A.~P.} 1979, Radiative Processes in Astrophysics (New York: Wiley)

\bibitem[Seward \& Wang(1988)]{Seward88}
Seward, F.~D., \& Wang, Z.-R.\ 1988, \apj, 332, 199 

\bibitem[Slane et al.(2000)]{Slane00}
Slane, P., Chen, Y., Schulz, N.~S., Seward, F.~D., Hughes, J.~P., \& Gaensler, B.~M.\ 2000, \apjl, 533, L29

\bibitem[Slane et al.(2008)]{Slane08} 
Slane, P., Helfand, D.~J., Reynolds, S.~P., Gaensler, B.~M., Lemiere, A., 
\& Wang, Z.\ 2008, \apjl, 676, L33 

\bibitem[\protect\citeauthoryear{{Trimble}, V.}{1973}]{Trimble73}
{{Trimble}, V.} 1973, \pasp, 85, 579

%\bibitem[\protect\citeauthoryear{Weekes et al.}{1989}]{Weekes89}
%{{Weekes}, T.~C., et al.} 1989, \apj, 342, 379

\bibitem[Willingale et al.(2001)]{Willingale01}
Willingale, R., Aschenbach, B., Griffiths, R.~G., Sembay, S., Warwick, R.~S., Becker, W., Abbey, A.~F., \& Bonnet-Bidaud, J.-M.\ 2001, \aap, 365, L212 

\end{thebibliography}

\end{document}